\documentstyle[amssymb,aps,prd, preprint]{revtex}

\begin{document}
\title{Remarks on some quantum effects in the spacetime of a cosmic string}
\author{Geusa de A. Marques$^{1,2}${\thanks{%
gmarques@fisica.ufpb.br}} and Valdir B. Bezerra$^{1}${\thanks{%
valdir@fisica.ufpb.br}}.}
\maketitle

\begin{center}
$^{1.}${\it {Departamento de F\'{\i}sica, Universidade Federal da
Para\'{\i}ba,}}

{\it {Caixa Postal 5008, Jo\~{a}o Pessoa, Pb, Brazil.}}

$^{2.}${\it {Departamento de F\'{\i}sica, Universidade Estadual da
Para\'{\i}ba,}}

{\it {Av. Juv\^{e}ncio Arruda S/N, Campina Grande, Pb, Brazil}}
\end{center}

\abstract

We study the behaviour of a non-relativistic quantum \ particle interacting
with different potentials, in the background spacetime generated by a cosmic
string. We find the energy spectra for the quantum systems under
consideration and discuss how they differ from their flat Minkowski
spacetime values.


\vskip3.0 cm \centerline{\bf{I. INTRODUCTION} }

The study concerning the influence of potentially observable effects of
gravitational fields at the atomic level has been an exciting research
field. Along this line of research the hydrogen atom, for example, has been
studied in the Friedmann-Robertson-Walker\cite{delta1} and \ Schwarzschild
spacetimes\cite{delta2} and also when it is placed in the gravitational
fields of a cosmic string and a global monopole\cite{delta3}. These studies
showed that an atom placed in a gravitational field is influenced by its
interaction with the local curvature as well as with the topology of the
spacetimes, and as a consequence there is a shift in the energy of each
atomic level which depends on the features of the spacetimes. The problem of
finding these shifts\cite{delta4} in the energy levels under the influence
of gravitational fields is of considerable theoretical as well as
observational interest.

General relativity, as a metric theory, predicts that gravitation is
manifested as the curvature of spacetime which is characterized by the
Riemann curvature tensor. Therefore, it seems interesting to know how the
curvature of spacetime at the position of the atom affects its spectrum. On
the other hand, we know that there are connections between topological
properties of the space and its local intrinsic geometry, and therefore it
is not possible to describe completely the physics of a given system based
only on its geometrical characteristics. Thus, it is also important to
investigate the role played by the topology on a given physical system.
Thus, the problem of finding how the energy spectrum of an atom is perturbed
by a gravitational field has to take into account the geometrical and
topological features of the considered spacetime as well. In other words,
the dynamic of atomic systems is determined by the curvature at the position
of the atom as well as by the topology of the background spacetime.

As examples of the influence of the topology on an atomic system, we can
mention the topological scattering in the context of quantum mechanics on a
cone$\cite{delta5},$ the interaction of a quantum system with conical
singularities \cite{delta6,delta7} and the modification of the energy levels
of a hydrogen atom\cite{delta3} placed in the gravitational fields of a
cosmic string\cite{delta8} and a global monopole\cite{delta9}.

The spacetime of a cosmic string is quite remarkable: its geometry is flat
everywhere apart from the symmetry axis. The spacetime around a cosmic
string is locally flat but not globally. Thus, the external gravitational
field due to a cosmic string may be approximately described by a commonly
called conical geometry. As a consequence a particle placed at rest around a
straight, infinite, static string will no be attracted to it. Therefore,
there is no local gravity in the space surrounding a cosmic string. On the
other hand, this conical structure can induce several effects like, for
example, gravitational lensing\cite{delta10}, electrostatic self-force on a
charge at rest\cite{delta11}, or on a line of charge\cite{delta12}, shifts
in the energy levels of a hydrogen atom\cite{delta3} and the gravitational
analogue \cite{delta13} of the well known electromagnetic Aharonov-Bohm
effect\cite{delta14}.

This paper deals with the problem concerning the effects of the nontrivial
topology of the spacetime generated by a cosmic string at the atomic level.
In order to investigate this, we address the question of how the shifts in
the energy spectrum of a particle are when it experiences different
potentials, like the generalized Kratzer\cite{delta15} plus a delta
potential, and Morse\cite{delta15} plus a delta potential. The influence of
the conical geometry on the energy spectrum manifests itself as a kind of a
gravitational Aharonov-Bohm effect for bound states\cite
{delta6,delta7,delta16}, whose analogue in the electromagnetic case shows
that\cite{delta17} the bound state energy depends on the magnetic flux
through a region from which the electron is excluded.

To begin with these studies, let us first solve the Schr\"{o}dinger equation
which reads
\[
i\hbar \frac{\partial \Psi }{\partial t}=-\frac{\hbar ^{2}}{2\mu }\nabla
_{LB}^{2}\Psi +V\Psi
\]
where $\nabla _{LB}^{2}=g^{-1/2}\partial _{i}\left( g^{ij}g^{1/2}\partial
_{j}\right) $, with $i,j=1,2,3;$ $g=\det \left( g_{ij}\right) $, is the
Laplace-Beltrami operator in the spacetime geometry surrounding a cosmic
string, \ and $V$ is the potential experienced by the particle.

This paper is organized as follows. In section 2, we discuss the energy
shifts of a particle placed in the gravitational field of a cosmic string,
which experiences a generalized Kratzer plus a delta function potentials. In
section 3, we consider Morse potential plus a delta function potential in
the same background. Section 4, is devoted to present our conclusions.

\newpage

\centerline{\bf {II. KRATZER AND DELTA FUNCTION INTERACTIONS}} %
\centerline{\bf { IN THE SPACETIME OF A COSMIC STRING}}

In what follows we will determine the energy spectrum of a particle which
experinces a generalized Kratzer potential plus a delta function potential
and we will analyse the role played by the cosmic string on this atomic
system. This result will be compared with the one obtained in the case
considered in the literature\cite{delta15} corresponding to the Kratzer
potential in a flat background. To do this let us consider the exterior
metric of an infinitely long straight and static string in spherical
coordinates, which is written as
\begin{equation}
ds^{2}=-dt^{2}+dr^{2}+r^{2}d\theta ^{2}+\alpha ^{2}r^{2}\sin ^{2}\theta
d\varphi ^{2},  \label{d1}
\end{equation}
with $0<r<\infty ,$ $0<\theta <\pi $, $0<\varphi \leq 2\pi $. The parameter $%
\alpha =1-4G\bar{\mu}$ runs in the interval $(0,1]$, $\bar{\mu}$ being the
linear mass density of the cosmic string (In this paper we will consider $%
c=1 $). Note that, in the special case $\alpha =1,$ Eq. (\ref{d1})
corresponds to the Minkowski spacetime in spherical coordinates.

The time-independent Schr\"{o}dinger equation in this background is given by
\begin{eqnarray}
&&-\frac{\hbar ^{2}}{2\mu r^{2}}\left[ \frac{\partial }{\partial r}\left(
r^{2}\frac{\partial }{\partial r}\right) +\cot \theta \frac{\partial }{%
\partial \theta }+\frac{\partial ^{2}}{\partial \theta ^{2}}+\frac{1}{\alpha
^{2}\sin ^{2}\theta }\frac{\partial ^{2}}{\partial \varphi ^{2}}\right] \Psi
\left( r,\theta ,\varphi \right)  \nonumber \\
&&+V(r)\Psi \left( r,\theta ,\varphi \right) \left. =\right. E\Psi \left(
r,\theta ,\varphi \right) .  \label{a}
\end{eqnarray}

Substituting the expression for the generalized Kratzer potential $%
V(r)=-2D\left( \frac{A}{r}-\frac{1}{2}\frac{B}{r^{2}}\right) ,$ with $D,$ $A$
and $B$ being positive constants (For $B=A^{2}$ we have the usual Kratzer
potential\cite{delta15}) and adding the delta function potential $T\delta
(r),$ where $T$ is a parameter which corresponds to the intensity of this
potential , Eq.(\ref{a}) turns into

\begin{equation}
\left[ \nabla _{LB}^{2}-k^{2}+\frac{4\mu }{\hbar ^{2}}D\left( \frac{A}{r}-%
\frac{1}{2}\frac{B}{r^{2}}\right) \right] \Psi (\vec{r})=\frac{2\mu }{\hbar
^{2}}T\delta (r)\Psi (\vec{r})  \label{3e}
\end{equation}
where

\begin{equation}
\nabla _{LB}^{2}=\frac{1}{r^{2}}\left[ \frac{\partial }{\partial r}\left(
r^{2}\frac{\partial }{\partial r}\right) +\cot \theta \frac{\partial }{%
\partial \theta }+\frac{\partial ^{2}}{\partial \theta ^{2}}+\frac{1}{\alpha
^{2}\sin ^{2}\theta }\frac{\partial ^{2}}{\partial \varphi ^{2}}\right] ,
\label{tt}
\end{equation}
and we are considering $k^{2}=-\frac{2\mu }{\hbar }E,$ with $E<0$ because we
are looking at the bound states.

Now, let us define $\bar{\nabla}_{LB}^{2}$ as \ \ \
\begin{equation}
\bar{\nabla}^{2}\equiv \nabla _{LB}^{2}+\frac{4\mu }{\hbar ^{2}}D\left(
\frac{A}{r}-\frac{1}{2}\frac{B}{r^{2}}\right) ,  \label{b}
\end{equation}
and
\[
V_{\delta }(r)\equiv \frac{2\mu }{\hbar ^{2}}T\delta (r).
\]

Considering the above definitions, Eq. ($\ref{3e}$) reads
\begin{equation}
\left( \bar{\nabla}^{2}-k^{2}\right) \Psi (r,\theta ,\varphi )=V_{\delta
}(r)\Psi (r,\theta ,\varphi ).  \label{4e}
\end{equation}

\bigskip

We can always write the general solution of the Eq. (\ref{4e}) in the
following form
\begin{equation}
\Psi (\vec{r})=\Phi _{k}\left( \vec{r}\right) +\int_{-\infty }^{+\infty
}G_{km_{\left( \alpha \right) }}^{l_{\left( \alpha \right) }}\left( \vec{r}-%
\vec{r}^{\prime }\right) V_{\delta }(r^{\prime })\Psi (\vec{r}^{\prime })d%
\vec{r}^{\prime },  \label{5e}
\end{equation}
where $G_{km_{\left( \alpha \right) }}^{l_{\left( \alpha \right) }}\left(
\vec{r}-\vec{r}^{\prime }\right) $ is the Green$^{\prime }$s function \ of a
particle which experiences the generalized the Kratzer potential in the
spacetime of a cosmic string. The parameters $m_{(\alpha )}$ and $l_{(\alpha
)}$ are defined by\cite{geusa}
\[
m_{(\alpha )}=\frac{m}{\alpha };\text{ }l_{(\alpha )}=l-\left( 1-\frac{1}{%
\alpha }\right) m,
\]
where $l$ is the orbital angular momentum quantum number and $m$ is the
magnetic quantum number. The Green$^{\prime }$s function obeys the following
equation
\begin{equation}
\left( \bar{\nabla}^{2}-k^{2}\right) G_{km_{\left( \alpha \right)
}}^{l_{\left( \alpha \right) }}\left( \vec{r}-\vec{r}^{\text{ }\prime
}\right) =\delta \left( \vec{r}-\vec{r}^{\text{ }\prime }\right) ,
\label{6e}
\end{equation}
and the function $\Phi _{k}\left( \vec{r}\right) $ is a solution of the so
called homogeneous equation. Therefore, to determine $\Phi _{k}\left( \vec{r}%
\right) $ the equation to be solved is
\begin{equation}
\left( \bar{\nabla}^{2}-k^{2}\right) \Phi _{k}\left( \vec{r}\right) =0\text{.%
}  \label{7e}
\end{equation}
The solution of Eq. (\ref{7e}) is given by$\cite{geusa}$%
\begin{equation}
\Phi _{k}\left( \vec{r}\right) =C_{k}\text{ }Y_{l\left( \alpha \right)
}^{m\left( \alpha \right) }\left( \theta ,\varphi \right) _{1}F_{1}\left(
\lambda _{\left( \alpha \right) }-\frac{2DA\mu }{k\hbar ^{2}},2\lambda
_{\left( \alpha \right) };2kr\right) r^{\lambda _{\left( \alpha \right)
}-1}e^{-kr},  \label{8e}
\end{equation}
where
\[
\lambda _{\left( \alpha \right) }\equiv \frac{1}{2}+\frac{1}{2}\sqrt{%
1+4l_{(\alpha )}(l_{(\alpha )}+1)+8\frac{\mu BD}{\hbar ^{2}}},
\]
$C_{k}$ is a normalization constant, $Y_{l\left( \alpha \right) }^{m\left(
\alpha \right) }\left( \theta ,\varphi \right) $ are generalized spherical
harmonics in the sense that $m_{(\alpha )}$ and $l_{(\alpha )}$are not
necessarily integers and $_{1}F_{1}$ is the confluent hypergeometric
function.

Now, let us assume that we can write the Green function as

\begin{equation}
G_{km_{\left( \alpha \right) }}^{l_{\left( \alpha \right) }}\left( \vec{r}-%
\vec{r}^{\prime }\right) =\sum_{l_{\left( \alpha \right) }=0}^{\infty
}\sum_{m_{\left( \alpha \right) }=-l_{\left( \alpha \right) }}^{+l_{\left(
\alpha \right) }}g_{k}^{l_{\left( \alpha \right) }}\left( r,r^{\prime
}\right) Y_{l\left( \alpha \right) }^{m\left( \alpha \right) \ast }\left(
\theta ^{\prime },\varphi ^{\prime }\right) Y_{l\left( \alpha \right)
}^{m\left( \alpha \right) }\left( \theta ,\varphi \right) .  \label{9e}
\end{equation}

We can also write the delta function according the representation as follows
\begin{equation}
\delta \left( \vec{r}-\vec{r}^{\prime }\right) =\frac{\left( r-r^{\prime
}\right) }{\alpha r^{2}}\sum_{l_{\left( \alpha \right) }=0}^{\infty
}\sum_{m_{\left( \alpha \right) }=-l_{\left( \alpha \right) }}^{+l_{\left(
\alpha \right) }}Y_{l\left( \alpha \right) }^{m\left( \alpha \right) \ast
}\left( \theta ^{\prime },\varphi ^{\prime }\right) Y_{l\left( \alpha
\right) }^{m\left( \alpha \right) }\left( \theta ,\varphi \right) ,
\label{10e}
\end{equation}
where we used the addition \ theorem for spherical harmonics.

Then, putting Eqs. (\ref{9e}) and (\ref{10e}) into (\ref{6e}), we find
\begin{eqnarray}
&&\left( \bar{\nabla}^{2}-k^{2}\right) \sum_{l_{\left( \alpha \right)
}=0}^{\infty }\sum_{m_{\left( \alpha \right) }=-l_{\left( \alpha \right)
}}^{+l_{\left( \alpha \right) }}g_{k}^{l_{\left( \alpha \right) }}\left(
r,r^{\prime }\right) Y_{l_{\left( \alpha \right) }}^{m_{\left( \alpha
\right) }\ast }\left( \theta ^{\prime },\varphi ^{\prime }\right)
Y_{l_{\left( \alpha \right) }}^{m_{\left( \alpha \right) }}\left( \theta
,\varphi \right)  \nonumber \\
&=&\frac{\left( r-r^{\prime }\right) }{\alpha r^{2}}\sum_{l_{\left( \alpha
\right) }=0}^{\infty }\sum_{m_{\left( \alpha \right) }=-l_{\left( \alpha
\right) }}^{+l_{\left( \alpha \right) }}Y_{l_{\left( \alpha \right)
}}^{m_{\left( \alpha \right) }\ast }\left( \theta ^{\prime },\varphi
^{\prime }\right) Y_{l_{\left( \alpha \right) }}^{m_{\left( \alpha \right)
}}\left( \theta ,\varphi \right) .  \label{13e}
\end{eqnarray}

Using Eqs. (\ref{a}) and (\ref{b}), we conclude that $g_{k}^{l_{\left( \alpha
\right) }}\left( r,r^{\prime }\right) $ obeys the following equation
\begin{eqnarray}
&&\left\{ \frac{1}{r^{2}}\left[ \frac{d}{dr}\left( r^{2}\frac{d}{dr}\right) -%
\frac{l_{\left( \alpha \right) }}{r^{2}}\left( l_{\left( \alpha \right)
}+1\right) \right] \right.  \nonumber \\
&&\left. +\frac{4\mu D}{\hbar ^{2}}\left( \frac{A}{r}-\frac{B}{r^{2}}\right)
-k^{2}\right\} g_{k}^{l_{\left( \alpha \right) }}\left( r,r^{\prime }\right)
\left. =\right. \frac{\delta \left( r-r^{\prime }\right) }{\alpha r^{2}},
\label{14e}
\end{eqnarray}
where the fact that $\vec{L}^{2}Y_{l_{\left( \alpha \right) }}^{m_{\left(
\alpha \right) }}\left( \theta ,\varphi \right) =l_{\left( \alpha \right)
}\left( l_{\left( \alpha \right) }+1\right) \hbar ^{2}Y_{l_{\left( \alpha
\right) }}^{m_{\left( \alpha \right) }}\left( \theta ,\varphi \right) $ was
taken into account.

Now, interchanging the function $g_{k}^{l_{\left( \alpha \right) }}\left(
r,r^{\prime }\right) $ by $R(r)$ in accordance with the definition
\begin{equation}
g_{k}^{l_{\left( \alpha \right) }}\left( r,r^{\prime }\right) =\frac{R(r)}{r}%
,  \label{c}
\end{equation}
we find that

\begin{equation}
\frac{d^{2}R}{dr^{2}}+\frac{4\mu DA}{r\hbar ^{2}}R+\left[ \frac{\frac{1}{4}%
-P^{2}}{r^{2}}-k^{2}\right] R=\frac{\delta \left( r-r^{\prime }\right) }{%
\alpha r},  \label{16e}
\end{equation}
where
\[
P\equiv \sqrt{l_{\left( \alpha \right) }\left( l_{\left( \alpha \right)
}+1\right) +\frac{2B\mu D}{\hbar ^{2}}+\frac{1}{4}}\geqslant \frac{1}{2}.
\]

By introducing the new variable $\rho $ defined by the relation
\[
\rho =2kr,
\]
Eq. (\ref{16e}) becomes
\begin{equation}
\frac{d^{2}R}{d\rho ^{2}}+\frac{\sigma }{\rho }R+\left[ \frac{\frac{1}{4}%
-P^{2}}{\rho ^{2}}-\frac{1}{4}\right] R=\frac{\delta \left( \rho -\rho
^{\prime }\right) }{\alpha \rho },  \label{17e}
\end{equation}
with the following definition
\[
\sigma \equiv \frac{4\mu DA}{2k\hbar ^{2}}.
\]

The solutions of \ the above differential equation can be obtained by the
standard procedure. When $\rho \neq \rho ^{\prime },$ the solution of Eq. (%
\ref{17e}) is given by
\begin{equation}
R(\rho )=B_{\sigma ,P}M_{\sigma ,P}\left( \rho \right) +C_{\sigma
.P}W_{\sigma ,P}\left( \rho \right) ,  \label{18e}
\end{equation}
where \ $M_{\sigma ,P}\left( \rho \right) $ and $W_{\sigma ,P}\left( \rho
\right) $ are \ the Whittaker functions.

In the region $\rho <\rho ^{\prime }$, we get
\begin{equation}
R\left( \rho _{<}\right) =B_{\sigma ,P}M_{\sigma ,P}\left( \rho \right) .
\label{19e}
\end{equation}

In the region $\rho >\rho ^{\prime }$, we have
\begin{equation}
R\left( \rho _{>}\right) =C_{\sigma ,P}W_{\sigma ,P}\left( \rho \right) ,
\label{20e}
\end{equation}
where $\rho _{<}$ $\left( \rho _{>}\right) $ represents the smaller
(greater) between $\rho $ and $\rho ^{\prime }.$

We can determine the constants \ $B_{\sigma ,P}$ and $C_{\sigma ,P}$ using
the continuity of the function $R(\rho )$ at $\rho =\rho ^{\prime }$ and the
discontinuity of its first derivative at this point. The continuity of $%
R\left( \rho \right) $ at $\rho =\rho ^{\prime }$ gives
\begin{equation}
B_{\sigma ,P}M_{\sigma ,P}\left( \rho ^{\prime }\right) =C_{\sigma
,P}W_{\sigma ,P}\left( \rho ^{\prime }\right) .  \label{21e}
\end{equation}

Integrating both sides of Eq. (\ref{17e}) in the interval $\rho ^{\prime
}-\varepsilon $ and $\rho ^{\prime }+\varepsilon $ with $\varepsilon
\rightarrow 0,$ we get the discontinuity of $R\left( \rho \right) $ $\ $at $%
\rho =\rho ^{\prime }$, which is given by
\begin{equation}
\left[ \frac{dR\left( \rho _{>}\right) }{d\rho }-\frac{dR\left( \rho
_{<}\right) }{d\rho }\right] _{\rho =\rho ^{\prime }}=\frac{1}{\alpha \rho
^{\prime }}.  \label{22e}
\end{equation}
Thus, substituting Eqs. (\ref{19e}) and (\ref{20e}) into (\ref{22e}), and
using Eq. (\ref{21e}), we \ get
\begin{equation}
C_{\sigma ,P}\tilde{W}\left| _{\rho =\rho ^{\prime }}\right. =\frac{%
M_{\sigma ,P}(\rho ^{\prime })}{\alpha \rho ^{\prime }},  \label{23e}
\end{equation}
with
\begin{equation}
\tilde{W}\left( \rho \right) =M_{\sigma ,P}(\rho )\frac{d}{d\rho }W_{\sigma
,P}(\rho )-W_{\sigma ,P}(\rho )\frac{d}{d\rho }M_{\sigma ,P}(\rho )
\label{24e}
\end{equation}
being the Wronskian of \ $M_{\sigma ,P}(\rho )$ and $W_{\sigma ,P}(\rho )$.

In order to compute $\tilde{W}$, we shall use the asymptotic expressions for
$M_{\sigma ,P}(\rho )$ and $W_{\sigma ,P}(\rho ).$ These functions behave,
near the origin, as \cite{Ma}
\begin{equation}
M_{\sigma ,P}(\rho )\rightarrow \rho ^{\frac{1}{2}+P}  \label{25e}
\end{equation}
and
\begin{equation}
W_{\sigma ,P}(\rho )\rightarrow \frac{\Gamma \left( 2P\right) }{\Gamma
\left( \frac{1}{2}+P-\sigma \right) }\rho ^{\frac{1}{2}-P}.  \label{26a}
\end{equation}
Using these equations we may infer that, the Wronskian reads
\begin{equation}
\tilde{W}=-2P\frac{\Gamma \left( 2P\right) }{\Gamma \left( \frac{1}{2}%
+P-\sigma \right) }.  \label{27e}
\end{equation}
Finally, substituting Eq. (\ref{27e}) into Eq. $\left( \ref{23e}\right) ,$
we obtain
\begin{equation}
C_{\sigma ,P}=-\frac{\Gamma \left( \frac{1}{2}+P-\sigma \right) }{2P\Gamma
\left( 2P\right) }\frac{M_{\sigma ,P}(\rho ^{\prime })}{\alpha \rho ^{\prime
}},  \label{28e}
\end{equation}
and using Eq. (\ref{24e}), we get
\begin{equation}
B_{\sigma ,P}=-\frac{\Gamma \left( \frac{1}{2}+P-\sigma \right) }{2P\Gamma
\left( 2P\right) }\frac{W_{\sigma ,P}(\rho ^{\prime })}{\alpha \rho ^{\prime
}}.  \label{29e}
\end{equation}
Substituting Eq. (\ref{28e}) and $\left( \ref{29e}\right) $ into (\ref{18e}),
we find that

\begin{equation}
R(r)=-\frac{\Gamma \left( \frac{1}{2}+P-\sigma \right) }{2P\Gamma \left(
2P\right) }\left[ \frac{W_{\sigma ,P}(2kr^{\text{ }\prime })}{2k\alpha r^{%
\text{ }\prime }}M_{\sigma ,P}(2kr)+\frac{M_{\sigma ,P}(2kr^{\text{ }\prime
})}{2k\alpha r^{\text{ }\prime }}W_{\sigma ,P}(2kr)\right] ,  \label{31e}
\end{equation}
where we have returned to the original variable.

Now, we can put Eq. (\ref{31e}) into (\ref{c}) and then substitute this
obtained result into Eq. (\ref{9e}) to find the following expression for
the Green$^{\prime }$s function
\begin{eqnarray}
G_{km_{\left( \alpha \right) }}^{l_{\left( \alpha \right) }}\left( \vec{r}-%
\vec{r}^{\text{ }\prime }\right) &=&-\sum_{l_{\left( \alpha \right)
}=0}^{\infty }\sum_{m_{\left( \alpha \right) }=-l_{\left( \alpha \right)
}}^{+l_{\left( \alpha \right) }}\frac{\Gamma \left( \frac{1}{2}+P-\sigma
\right) }{2P\Gamma \left( 2P\right) }\left[ \frac{W_{\sigma ,P}(2kr^{\text{ }%
\prime })}{2k\alpha r^{\text{ }\prime }r}M_{\sigma ,P}(2kr)\right.  \nonumber
\\
&&\left. +\frac{M_{\sigma ,P}(2kr^{\text{ }\prime })}{2k\alpha r^{\text{ }%
\prime }r}W_{\sigma ,P}(2kr)\right] Y_{l\left( \alpha \right) }^{m\left(
\alpha \right) \ast }\left( \theta ^{\prime },\varphi ^{\prime }\right)
Y_{l\left( \alpha \right) }^{m\left( \alpha \right) }\left( \theta ,\varphi
\right) .  \label{33e}
\end{eqnarray}

Therefore, the radial part of solution of the Eq. (\ref{4e}) reads

\begin{eqnarray}
\Psi _{km_{\left( \alpha \right) }}^{l_{\left( \alpha \right) }}(r) &=&\Phi
_{k}\left( r\right) +\frac{2\mu T}{\hbar ^{2}}\int_{-\infty }^{+\infty
}G_{km_{\left( \alpha \right) }}^{l_{\left( \alpha \right) }}\left( r-r^{%
\text{ }\prime }\right) \delta (r^{\text{ }\prime })\Psi (r^{\text{ }\prime
})dr^{\text{ }\prime }  \nonumber \\
&=&C_{k}\text{ }_{1}F_{1}\left( \lambda _{\left( \alpha \right) }-\frac{%
2DA\mu }{k\hbar ^{2}},2\lambda _{\left( \alpha \right) };2kr\right)
r^{\lambda _{\left( \alpha \right) }-1}e^{ikr}+I_{1}+I_{2},  \label{34e}
\end{eqnarray}
where
\[
I_{1}=-\frac{2\mu T}{\hbar ^{2}}\frac{\Gamma \left( \frac{1}{2}+P-\sigma
\right) }{2P\Gamma \left( 2P\right) }\int_{-\infty }^{r^{\prime }}\frac{%
W_{\sigma ,P}(2kr^{\text{ }\prime \prime })}{2k\alpha r^{\text{ }\prime
\prime }r}M_{\sigma ,P}(2kr)\delta (r^{\text{ }\prime \prime })\Psi (r^{%
\text{ }\prime \prime })dr^{\text{ }\prime \prime },
\]
and
\[
I_{2}=-\frac{2\mu T}{\hbar ^{2}}\frac{\Gamma \left( \frac{1}{2}+P-\sigma
\right) }{2P\Gamma \left( 2P\right) }\int_{r^{\prime }}^{\infty }\frac{%
M_{\sigma ,P}(2kr^{\text{ }\prime \prime })}{2k\alpha r^{\text{ }\prime
\prime }r}W_{\sigma ,P}(2kr)\delta (r^{\text{ }\prime \prime })\Psi (r^{%
\text{ }\prime \prime })dr^{\text{ }\prime \prime }.
\]

Let us analyse the terms given by $I_{1}$ and $I_{2}$ in two different
cases, namely, $P>\frac{1}{2}$ which means $\lambda _{\left( \alpha \right)
}>1;$ and $P=\frac{1}{2}$ which implies that $\lambda _{\left( \alpha
\right) }=1.$ In the former case we have

\begin{equation}
\lim_{r^{\text{ }\prime \prime }\rightarrow 0}\frac{W_{\sigma ,P}}{r^{\text{
}\prime \prime }}=\lim_{r^{\text{ }\prime \prime }\rightarrow 0}\left( \frac{%
\Gamma \left( 2P\right) }{\Gamma \left( \frac{1}{2}+P-\sigma \right) }\frac{%
\left( 2kr^{\text{ }\prime \prime }\right) ^{\frac{1}{2}-P}}{r^{\text{ }%
\prime \prime }}\right) \rightarrow \infty ,  \label{35e}
\end{equation}
due to the fact that $\Psi (r)$ must be a well behaved function. This result
implies that $M_{\sigma ,P}\left( 2kr^{\text{ }\prime \prime }\right) =0.$
We also have that
\begin{equation}
\lim_{r^{\text{ }\prime \prime }\rightarrow 0}\frac{M_{\sigma ,P}}{r^{\text{
}\prime \prime }}=\lim_{r^{\text{ }\prime \prime }\rightarrow 0}\left( \frac{%
\left( 2kr^{\text{ }\prime \prime }\right) ^{\frac{1}{2}+P}}{r^{\text{ }%
\prime \prime }}\right) \rightarrow 0.  \label{36e}
\end{equation}
From these results we can conclude that

\[
I_{1}=I_{2}=0,
\]
and therefore, we have
\begin{equation}
\Psi (r)=C_{k1}F_{1}\left( \lambda _{\left( \alpha \right) }-\frac{2\mu DA}{%
\hbar ^{2}k},2\lambda _{\left( \alpha \right) };2kr\right) e^{-kr}r^{\lambda
_{\left( \alpha \right) }-1}.  \label{37e}
\end{equation}
This result means that in this case there is no contribution coming from the
delta function interaction. In order to avoid the divergence of this
solution, the following condition must hold
\[
\lambda _{\left( \alpha \right) }-\frac{\nu ^{2}}{Ak}=-n;\text{ }%
n=0,1,2,...,
\]
where
\[
\nu ^{2}=\frac{2\mu DA^{2}}{\hbar ^{2}}.
\]
From the above condition we find that the energy spectrum is given by
\begin{equation}
E=-\frac{\hbar ^{2}}{2\mu A^{2}}\nu ^{4}\left\{ n+\frac{1}{2}+\sqrt{\frac{1}{%
4}+l_{\left( \alpha \right) }\left( l_{\left( \alpha \right) }+1\right) +%
\frac{2\mu DB}{\hbar ^{2}}}\right\} ^{-2}.  \label{38e}
\end{equation}
Therefore, in this case, the energy spectrum for this problem is the same
obtained\cite{geusa} for a particle in the presence of a Coulomb potntial
and in the absence of a delta function interaction. This result means that the
presence of this function does not influence the energy spectrum.

In the second case, that is, $P=\frac{1}{2},$ we have just the $S$ wave,
because in this situation \ the only possible value of $l_{\left( \alpha
\right) }$ is zero. We also have necessarily that $B=0$, in order to fulfill
this condition. Then, we have

\begin{equation}
\Psi _{k}(r)=C_{k1}F_{1}\left( 1-\frac{i\nu ^{2}}{2kA},2;2kr\right) e^{-kr}+%
\frac{T}{\alpha }C_{k1}^{\prime }F_{1}\left( -\frac{\nu ^{2}}{2kA}%
,0;2kr\right) .  \label{39e}
\end{equation}
In order to get a physically acceptable solution, we must impose the
following conditions

i)
\[
1-\frac{i\nu ^{2}}{2kA}=-n;\text{ }n=0,1,2...,
\]
and

ii)
\[
-\frac{\nu ^{2}}{2kA}=-n^{\prime };\text{ }n^{\prime }=0,1,2,...\text{ .}
\]
These conditions are compatible only for $n=1$ and $n^{\prime }=0$, which
means that \ the energy diverges. This result tell us that in this situation
the delta function prevents the existence of bound states.

\vskip1.0 cm

\centerline{\bf {II. MORSE AND DELTA FUNCTION INTERACTIONS}}
\centerline{\bf
{ IN THE SPACETIME OF A COSMIC STRING}}

In what follows we analyse the role played by the Morse and delta function
potentials on the energy spectrum of a nonrelativistic quantum particle
placed in the presence of a cosmic string.

The Morse potential\cite{delta15}, which is used to describe the vibrations
of a diatomic molecule reads
\begin{equation}
V(r)=D\left( e^{-2\tilde{\alpha}x}-2e^{-\tilde{\alpha}x}\right) ,  \label{1d}
\end{equation}
where $x=\frac{r-r_{0}}{r_{0}}$ and $\tilde{\alpha}$ is a positive real
number.

Considering this potential, Eq. (\ref{3e}) reads

\begin{equation}
\left[ \nabla _{LB}^{2}-k^{2}-\frac{2\mu }{\hbar ^{2}}D\left( e^{-2\tilde{%
\alpha}x}-2e^{-\tilde{\alpha}x}\right) \right] \Psi (\vec{r})=\frac{2\mu }{%
\hbar ^{2}}T\delta (r)\Psi (\vec{r}).  \label{42e}
\end{equation}

As pointed out in the previous section, the solution of Eq. (\ref{42e}) can
be given by Eq. (\ref{5e}), with $\Phi _{k}\left( \vec{r}\right) $ being the
solution of the homogeneous equation, Eq. (\ref{7e}), \ where $\bar{\nabla}%
^{2}$ is now written as
\begin{equation}
\bar{\nabla}^{2}\equiv \nabla _{LB}^{2}-\frac{2\mu }{\hbar ^{2}}D\left( e^{-2%
\tilde{\alpha}x}-2e^{-\tilde{\alpha}x}\right) .  \label{456}
\end{equation}
As the Morse potential depends only on the radial coordinate, we can
separate the solution $\Phi _{k}\left( \vec{r}\right) $ as
\begin{equation}
\Phi _{k}\left( \vec{r}\right) =\chi (r)Y_{l\left( \alpha \right) }^{m\left(
\alpha \right) }\left( \theta ,\varphi \right) .  \label{46e}
\end{equation}
Substituting Eq. (\ref{46e}) into the homogeneous equation for $\Phi
_{k}\left( \vec{r}\right) $, the equation to be solved for $\chi(r) $ is
\begin{equation}
\left[ 2r\frac{d}{dr}+r^{2}\frac{d^{2}}{dr^{2}}-l_{\left( \alpha \right)
}\left( l_{\left( \alpha \right) }+1\right) -\frac{2\mu }{\hbar ^{2}}D\left(
e^{-2\tilde{\alpha}x}-2e^{-\tilde{\alpha}x}\right) r^{2}-k^{2}r^{2}\right]
\chi (r)=0.  \label{47e}
\end{equation}
By making the substitution
\[
R(r)=r\chi (r)
\]
we get that $R(r)$ obeys the following equation
\begin{equation}
\frac{d^{2}R(x)}{dx^{2}}-r_{0}^{2}\left( \frac{l_{\left( \alpha \right) }}{%
\left( x+1\right) ^{2}}\left( l_{\left( \alpha \right) }+1\right) +\frac{%
2\mu D}{\hbar ^{2}}\left( e^{-2\tilde{\alpha}x}-2e^{-\tilde{\alpha}x}\right)
+k^{2}\right) R(x)=0,  \label{50e}
\end{equation}
in which we have reintroduced the variable $x=\frac{r-r_{0}}{r_{0}}$ and the
definition
\begin{equation}
\varsigma ^{2}\equiv \frac{2\mu Dr_{0}^{2}}{\hbar ^{2}}.  \label{49e}
\end{equation}

Expanding $\left( x+1\right) ^{-2}$ and considering $\left| x\right| \ll
1,\, $we find that

\begin{equation}
c_{1}e^{-\tilde{\alpha}x}+c_{2}e^{-2\tilde{\alpha}x}=\frac{3}{\tilde{\alpha}}%
-2x+\frac{2}{3}\tilde{\alpha}x^{3}-\frac{2}{\tilde{\alpha}^{2}}+3x^{2}-3%
\tilde{\alpha}x^{3}+...,  \label{52e}
\end{equation}
where
\[
c_{1}\equiv \frac{4}{\tilde{\alpha}}-\frac{6}{\tilde{\alpha}^{2}};\text{ and
\quad }c_{2}\equiv -\frac{1}{\tilde{\alpha}}+\frac{3}{\tilde{\alpha}^{2}}.
\]
Now, considering
\[
c_{0}\equiv 1-\frac{3}{\tilde{\alpha}}+\frac{3}{\tilde{\alpha}^{2}},
\]
we get
\begin{equation}
c_{0}+c_{1}e^{-\tilde{\alpha}x}+c_{2}e^{-2\tilde{\alpha}x}=1-2x+3x^{2}-%
\left( 3\tilde{\alpha}-\frac{2}{3}\tilde{\alpha}^{2}\right) x^{3}+...\text{ ,%
}  \label{53e}
\end{equation}
which corresponds, approximately to the expansion of $\left( x+1\right)
^{-2},$ that is
\begin{equation}
\left( x+1\right) ^{-2}\cong c_{0}+c_{1}e^{-\tilde{\alpha}x}+c_{2}e^{-2%
\tilde{\alpha}x}.  \label{54e}
\end{equation}

Substituting Eq. (\ref{54e}) into Eq. (\ref{50e}), we find
\begin{equation}
\frac{d^{2}R(x)}{dx^{2}}+\left[ -\beta ^{2}+2\gamma _{1}^{2}e^{-\tilde{\alpha%
}x}-\gamma _{2}^{2}e^{-2\tilde{\alpha}x}\right] R(x)=0,  \label{55e}
\end{equation}
where
\begin{eqnarray*}
\beta ^{2} &=&l_{\left( \alpha \right) }\left( l_{\left( \alpha \right)
}+1\right) c_{0}+r_{0}^{2}k^{2}, \\
\gamma _{1}^{2} &=&\varsigma ^{2}-\frac{c_{1}}{2}l_{\left( \alpha \right)
}\left( l_{\left( \alpha \right) }+1\right) , \\
\gamma _{2}^{2} &=&\varsigma ^{2}+c_{2}l_{\left( \alpha \right) }\left(
l_{\left( \alpha \right) }+1\right) .
\end{eqnarray*}

Now, let us define
\begin{equation}
y\equiv \xi e^{-\tilde{\alpha}x},  \label{56e}
\end{equation}
with
\[
\xi \equiv 2\frac{\gamma _{2}}{\alpha }.
\]

Thus, in terms of this new variable, Eq. (\ref{55e}) reads

\begin{equation}
y^{2}\frac{d^{2}R(y)}{dy^{2}}+y\frac{dR(y)}{dy}+\frac{1}{\tilde{\alpha}}%
\frac{\gamma _{1}^{2}}{\gamma _{2}}yR(y)+\left[ -\left( \frac{\beta }{\tilde{%
\alpha}}\right) ^{2}-\frac{y^{2}}{4}\right] R(y)=0.  \label{57e}
\end{equation}
This equation can be written as
\begin{equation}
\frac{d^{2}F(y)}{dy^{2}}+\frac{\tilde{\sigma}}{y}F(y)+\left[ \frac{\frac{1}{4%
}-\tilde{P}^{2}}{y^{2}}-\frac{1}{4}\right] F(y)=0,  \label{58e}
\end{equation}
where
\[
\tilde{\sigma}\equiv \frac{\gamma _{1}^{2}}{\gamma _{2}\tilde{\alpha}};\text{
}\tilde{P}^{2}\equiv \frac{\beta ^{2}}{\tilde{\alpha}^{2}}
\]
and we have made the substitution $R(y)\equiv \frac{1}{\sqrt{y}}F(y).$

Equation (\ref{58e}) is a Whittaker differential equation, whose regular
solution at the origin is
\begin{equation}
F(y)=e^{-\frac{1}{2}y}y^{\frac{1}{2}+\left| \frac{\beta }{\tilde{\alpha}}%
\right| }M\left( \frac{1}{2}+\left| \frac{\beta }{\tilde{\alpha}}\right| -%
\tilde{\sigma},1+2\left| \frac{\beta }{\tilde{\alpha}}\right| ;y\right) .
\label{59e}
\end{equation}

Therefore, the solution $\Phi \left( \vec{r}\right) ,$ can be written in
terms of the variables $r,\theta $ and $\varphi $ as

\begin{eqnarray}
\Phi \left( \vec{r}\right) &=&Y_{l_{\left( \alpha \right) }}^{m_{\left(
\alpha \right) }}\left( \theta ,\varphi \right) \frac{\exp \left( -\frac{1}{2%
}\xi e^{-\tilde{\alpha}\left( \frac{r}{r_{0}}-1\right) }\right) }{\sqrt{%
^{\xi }e^{-\tilde{\alpha}\left( \frac{r}{r_{0}}-1\right) }}}\left( \xi e^{-%
\tilde{\alpha}\left( \frac{r}{r_{0}}-1\right) }\right) ^{\frac{1}{2}+\left|
\frac{\beta }{\tilde{\alpha}}\right| }  \nonumber \\
&&\times M\left( \frac{1}{2}+\left| \frac{\beta }{\tilde{\alpha}}\right| -%
\frac{\gamma _{1}^{2}}{\gamma _{2}\tilde{\alpha}},1+2\left| \frac{\beta }{%
\tilde{\alpha}}\right| ;\xi e^{-\tilde{\alpha}\left( \frac{r}{r_{0}}%
-1\right) }\right) .  \label{61e}
\end{eqnarray}

Now, we want to determine the second part of the solution of Eq. (\ref{42e})
which corresponds to the second term in the right-hand side of Eq. (\ref{9e}%
) for the present case. In order to do this, we have to calculate the Green$%
^{\prime }$s function corresponding to the problem under consideration.

Proceeding in analogy with the previous section, we can determine the Green$%
^{\prime }$s function from the following equation

\begin{eqnarray*}
&&\left\{ \frac{1}{r^{2}}\left[ \frac{d}{dr}\left( r^{2}\frac{d}{dr}\right)
-l_{\left( \alpha \right) }\left( l_{\left( \alpha \right) }+1\right) \right]
\right. \\
&&\left. -\frac{2\mu D}{\hbar ^{2}}\left( e^{-2\tilde{\alpha}x}-2e^{-\tilde{%
\alpha}x}\right) -k^{2}\right\} g_{k}^{l_{\left( \alpha \right) }}\left(
r,r^{\prime }\right) \left. =\right. \frac{\delta \left( r-r^{\prime
}\right) }{\alpha r^{2}},
\end{eqnarray*}

\begin{eqnarray*}
&&\left\{ \left[ \frac{d}{dr}\left( r^{2}\frac{d}{dr}\right) -l_{\left(
\alpha \right) }\left( l_{\left( \alpha \right) }+1\right) \right] \right. \\
&&\left. -\frac{2\mu D}{\hbar ^{2}}\left( e^{-2\tilde{\alpha}x}-2e^{-\tilde{%
\alpha}x}\right) r^{2}-k^{2}r^{2}\right\} g_{k}^{l_{\left( \alpha \right)
}}\left( r,r^{\prime }\right) \left. =\right. \frac{\delta \left(
r-r^{\prime }\right) }{\alpha }.
\end{eqnarray*}
Doing the same transformations as before, we find that

\begin{equation}
\frac{d^{2}T(y)}{dy^{2}}+\frac{\sigma }{y}T(y)+\left[ \frac{\frac{1}{4}-%
\frac{\beta ^{2}}{\tilde{\alpha}}}{y^{2}}-\frac{1}{4}\right] T(y)=\frac{%
y^{\prime }\delta \left( y-y^{\prime }\right) }{y^{2}r_{0}^{2}\tilde{\alpha}%
\alpha \left( 1-\frac{1}{\tilde{\alpha}}\ln \left( \frac{y}{\zeta }\right)
\right) },  \label{65e}
\end{equation}
where $R(y)=\frac{1}{\sqrt{y}}T(y).$

To solve this equation, we have to consider two different regions, namely, $%
y<y^{\prime }$ and $y>y^{\prime }$. In these regions we have that $y\neq
y^{\prime }$, and as a consequence Eq. (\ref{65e}) can be written simply as
\begin{equation}
\frac{d^{2}T(y)}{dy^{2}}+\frac{\sigma }{y}T(y)+\left[ \frac{\frac{1}{4}-P^{2}%
}{y^{2}}-\frac{1}{4}\right] T(y)=0,  \label{66e}
\end{equation}
where
\[
P^{2}=\frac{\beta ^{2}}{\tilde{\alpha}}.
\]

The solution of Eq. (\ref{66e}) is
\begin{equation}
T(y)=B_{\sigma ,P}M_{\sigma ,P}\left( y\right) +C_{\sigma ,P}W_{\sigma
,P}\left( y\right) .  \label{67e}
\end{equation}
 With this obtained result for $T(y)$ and taking
into account the relation of this function with $g_{k}^{l_{\left( \alpha
\right) }}\left( r,r^{\prime }\right) $, \ which enter the expression of the
Green$^{\prime }$s function, we can write the following result bellow
\begin{eqnarray}
g_{k}^{l_{\left( \alpha \right) }}\left( r,r^{\prime }\right) &=&-\frac{%
\zeta ^{-\frac{1}{2}}}{r}\exp \left\{ \frac{\tilde{\alpha}}{2}\left( \frac{r%
}{r_{0}}-1\right) \frac{\Gamma \left( \frac{1}{2}+P-\sigma \right) }{%
2P\Gamma \left( 2P\right) }\right\}  \nonumber \\
&&\left[ \frac{W_{\sigma ,P}\left( \zeta \exp \left\{ -\tilde{\alpha}\left(
\frac{r^{\prime }}{r_{0}}-1\right) \right\} \right) M_{\sigma ,P}\left(
\zeta \exp \left\{ -\tilde{\alpha}\left( \frac{r}{r_{0}}-1\right) \right\}
\right) }{\alpha \zeta \exp \left\{ -\tilde{\alpha}\left( \frac{r^{\prime }}{%
r_{0}}-1\right) \right\} r_{0}\tilde{\alpha}r^{\prime }}\right.  \nonumber \\
&&+\left. \frac{M_{\sigma ,P}\left( \zeta \exp \left\{ -\tilde{\alpha}\left(
\frac{r^{\prime }}{r_{0}}-1\right) \right\} \right) W_{\sigma ,P}\left(
\zeta \exp \left\{ -\tilde{\alpha}\left( \frac{r}{r_{0}}-1\right) \right\}
\right) }{\alpha \zeta \exp \left\{ -\tilde{\alpha}\left( \frac{r^{\prime }}{%
r_{0}}-1\right) \right\} r_{0}\tilde{\alpha}r^{\prime }}\right] .
\label{69e}
\end{eqnarray}
Putting Eq. (\ref{69e}) into the expression for the Green function given by
Eq. (\ref{9e}) and then substituting this result into Eq. ($\ref{9e}$) and
realizing the integration, we conclude that this vanishes and therefore the $%
\delta \left( r\right) $ function potential does not affect the final
result. This means that the presence of the delta function potential does
not influence the behaviour of the particle, which is determined, in this
case, only by the Morse potential.

Therefore, in order to determine the energy spectrum associated with this
quantum system in the spacetime of a cosmic string, we have to consider just
the solution corresponding to the single presence of the Morse potential
which is given by Eq. (\ref{59e}). For the wavefunction to remain finite as $%
y\rightarrow \infty $, in such a way to have a physically acceptable
solution, the following condition must be fulfilled
\begin{equation}
\frac{1}{2}+\left| \frac{\beta }{\tilde{\alpha}}\right| -\tilde{\sigma}=-n;%
\text{ }n=0,1,2,...\text{ .}  \label{final}
\end{equation}
Thus, using this condition we find that the energy spectrum is given by
\begin{eqnarray}
E &\cong &-\frac{\hbar ^{2}}{2\mu r_{0}^{2}}\left\{ -\varsigma
^{2}+2\varsigma \tilde{\alpha}\left( n+\frac{1}{2}\right) -\tilde{\alpha}%
^{2}\left( n+\frac{1}{2}\right) ^{2}-\frac{9}{4\varsigma ^{2}}l_{\left(
\alpha \right) }^{2}\left( l_{\left( \alpha \right) }+1\right) ^{2}\left(
\frac{1}{\tilde{\alpha}}-\frac{1}{\tilde{\alpha}^{2}}\right) ^{2}\right.
\nonumber \\
&&\left. +l_{\left( \alpha \right) }\left( l_{\left( \alpha \right)
}+1\right) -\frac{3}{\varsigma }l_{\left( \alpha \right) }\left( l_{\left(
\alpha \right) }+1\right) \left( 1-\frac{1}{\tilde{\alpha}}\right) \left( n+%
\frac{1}{2}\right) \right\} ,  \label{60e}
\end{eqnarray}
in which we have neglected terms of order $\varsigma ^{-6}$ and up.

\vskip 2.0 cm

\section*{VI. CONCLUSIONS}

In the spacetime of a cosmic string we studied the behaviour of a particle
in the presence of the Kratzer and Morse potentials in addition to a delta
function interaction.

In the case of Kratzer plus a delta function interactions, if one consider $%
l\neq 0$, there are bound states with the same energy corresponding to the
case in which the delta interaction is absent. This means that there is no
contribution coming from the delta interaction. In the case in which $l=0$
and $B=0$, the presence of the delta interaction prevents the existence of
bound states.

The case corresponding to the Morse potential plus a delta function
interaction, presents bound states which are not affected by the presence of
the delta function interaction in any situation. In this way the
introduction of a delta function gives us no additional contribution to the
energy spectra compared with the case in which this term is not present.

It is worth commenting that in both cases the obtained results tell us that
the energy spectra are modified as compared to the flat spacetime Minkowski
result and these shifts are connected with the conical structure of the
spacetime generated by a cosmic string. In other words, these shifts in the
energies are due completely to the topological features of this spacetime.

The magnitude of the modifications in the spectra of the quantum mechanical
systems by the presence of the cosmic string is measurable, in principle. To
be more realistic, if we want to produce observable modifications at the
astrophysical scale, one needs a large number of particles in the states
that we have studied. We can also have the possibility of getting a strong
contribution to real spectral effects if the cosmic strings occur in large
number, like, for example, in superfluid vortices in neutron stars which are
expected to occur with densities of $10^{6}$ vortices per square centimeter
of stellar cross section.

To end up, let us comment that the study of a quantum system in a
non-trivial gravitational background, like in the spacetime of a cosmic
string considered in this paper, may shed some light on the problems of
combining quantum mechanics and general relativity.

\section*{Acknowledgments}

We acknowledge Conselho Nacional de Desenvolvimento Cient\'{\i}fico e
Tecnol\'{o}gico (CNPq) and Coordena\c{c}\~{a}o de Aperfei\c{c}oamento de
Pessoal de N\'{\i}vel Superior (CAPES) - Programa PROCAD, for partial
financial support.


\vskip2.0 cm

\begin{references}
\bibitem{delta1}  J. Audretsch and G. Sch\"{a}fer, Gen. Rel. Grav. {\bf 9},
243 (1978).

\bibitem{delta2}  L. Parker, Phys. Rev. Lett. {\bf 44}, 1559 (1980).

\bibitem{delta3}  Geusa de A. Marques and Valdir B. Bezerra, Phys. Rev. {\bf %
D66}, 105011 (2002).

\bibitem{delta4}  L. Parker and L. Pimentel, Phys. Rev. {\bf D44}, 3180
(1982).

\bibitem{delta5}  P. de Sousa Gerbert and R. Jackiw, Commun. Math. Phys.
{\bf 124}, 229 (1989).

\bibitem{geusa}  Geusa de A. Marques and V. B. Bezerra, Class. Quantum Grav.
{\bf 19}, 985 (2002).

\bibitem{delta6}  V. B. Bezerra, Class. Quantum Grav. {\bf 8}, 1939 (1991).

\bibitem{delta7}  Cl\'{a}udio Furtado and Fernando Moraes, J. Phys. {\bf A33}%
, 5513 (2000).

\bibitem{delta8}  T. W. B. Kibble, J. Phys. {\bf A9}, 1387 (1976); Ya. B.
Zel'dovich, Mon. Not. R. Astron. Soc. {\bf 192}, 663 (1980); A. Vilenkin,
Phys. Rev. Lett. {\bf 46}, 1169 (1981).

\bibitem{delta9}  Manuel Barriola and A. Vilenkin, Phys. Rev. Lett. {\bf 63}%
, 341 (1989).

\bibitem{delta10}  J. R. Gott III, Astrophys. J. {\bf 288}, 422 (1985).

\bibitem{delta11}  B. Linet, Phys. Rev. {\bf D33}, 1833 (1986).

\bibitem{delta12}  E. R. Bezerra de Mello, V. B. Bezerra, C. Furtado and F.
Moraes, Phys. Rev. {\bf D51}, 7140 (1995).

\bibitem{delta13}  V. B. Bezerra, Phys. Rev. {\bf D35}, 2031 (1987).

\bibitem{delta14}  Y. Aharonov and D. Bohm, Phys. Rev. {\bf 119}, 485 (1959).

\bibitem{delta15}  S. Fl\"{u}gge, {\it Pratical Quantum Mechanics},
Springer-Verlag (1971).

\bibitem{delta16}  V. B. Bezerra, J. Math. Phys. {\bf 30}, 2895 (1989).

\bibitem{delta17}  M. Peshkin, Phys. Rep. {\bf 80}, 375 (1981).

\bibitem{Ma}  M. Abramowitz and I. Stegun, {\it Handbook of Mathematical
Functions}, 9$^{th}$ Edition, Dover (1972).
\end{references}
\end{document}